\documentclass[global,twocolumn]{svjour}
\usepackage{graphics}
\usepackage{graphicx}
\usepackage{amsmath,amssymb,amstext}
\usepackage[numbers,sort&compress,square]{natbib}
\usepackage{bm} 
\usepackage[english]{babel}
\usepackage{epstopdf}
\newcommand{\bra}[1]{\left\langle \left. #1 \right. \right|}
\newcommand{\ket}[1]{\left| \left. #1 \right. \right\rangle}
\newcommand{\ketbra}[1]{\ket{#1}\bra{#1}}
\def\out{_{\rm out}}
\def\Ppara{P_{\Vert}}
\def\Psucc{P_{\rm succ}}
\begin{document}
\title{Entanglement Distillation using the Exchange Interaction}
\author{
Adrian Auer\inst{1}\thanks{These authors contributed equally to this work.},
Ren\'e Schwonnek\inst{2}\thanks{These authors contributed equally to this work.}, 
Christian Schoder\inst{1}, 
Lars Dammeier\inst{2},
Reinhard F. Werner\inst{2}, 
Guido Burkard\inst{1}}
\institute{Department of Physics, University of Konstanz, D-78457 Konstanz, Germany
\and Institut f\"ur Theoretische Physik, Leibniz Universit\"at, D-30167 Hannover, Germany
}
\mail{adrian.auer@uni-konstanz.de}
\maketitle
\begin{abstract}
A key ingredient of quantum repeaters is entanglement distillation, i.e.,~the generation of  high-fidelity entangled qubits from a larger set of pairs with lower fidelity. Here, we present entanglement distillation protocols based on qubit couplings that originate from exchange interaction. First, we make use of asymmetric bilateral two-qubit operations generated from anisotropic exchange interaction and show how to distill entanglement using two input pairs. We furthermore consider the case of three input pairs coupled through isotropic exchange. Here, we characterize a set of protocols which are optimizing the tradeoff between the fidelity increase and the probability of a successful run.
\end{abstract}
\section{Introduction}
\label{intro}
In a quantum communication (QC) network, the establishment of long-distance entanglement is indispensable to fully harness the advantages offered by quantum information processing \cite{Kimble:2008}, e.g.~perfectly secure long-distance quantum communication \cite{PhysRevLett.67.661}. For the distribution of maximally entangled states, one has to counteract decoherence processes due to the unavoidable interaction of entangled particles with their environment. A fundamental component of a QC network are therefore quantum repeaters \cite{PhysRevLett.81.5932,PhysRevA.59.169}, which enable the successive creation of near-maximal entanglement between distant network nodes.
Entanglement distillation, on the other hand, is a key part of quantum repeaters, and requires a functioning quantum memory \cite{Simon:2010}.

Spins in solid-state environments, such as single electrons in semiconductor quantum dots (QDs) \cite{doi:10.1146/annurev-conmatphys-030212-184248} or nitrogen-vacancy centers in diamond \cite{doi:10.1146/annurev-conmatphys-030212-184238}, show long coherence times ($\mu$s to ms) and offer flexible controlling mechanisms. However, the original protocols of entanglement distillation \cite{PhysRevLett.76.722,PhysRevLett.77.2818} are rather unpractical, e.g.~for spin qubits in QDs mentioned above, since an efficient implementation of the required controlled-\textsc{not} (\textsc{cnot}) gates is very demanding. In the case of Heisenberg exchange, it requires two two-qubit interaction pulses each leading to the $\sqrt{\textsc{swap}}$ gate, and additionally five single qubit rotations to construct a \textsc{cnot} gate \cite{PhysRevA.57.120}. However, single-spin rotations take on the order of 100 ns \cite{Nowack30112007} and are therefore slow compared to exchange-based two qubit operations. The $\sqrt{\textsc{swap}}$ gate, e.g.,~has been succesfully implemented in less than 200 ps \cite{Petta30092005}.

This circumstance motivates the work presented in this contribution, namely a careful study of entanglement distillation protocols using only the typical interaction between electrons in QDs, namely the exchange interaction \cite{PhysRevA.57.120,PhysRevB.59.2070}. In the following, we first extend an earlier proposal based on isotropic Heisenberg exchange \cite{PhysRevA.90.022320}, where the concept of asymmetric bilateral two-qubit operations for protocols using two input pairs was introduced, to the more general scenario of an anisotropic exchange interaction. 

Furthermore, we analyze entanglement distillation protocols for exchange-coupled qubits that use three input pairs. Our method is based on an algebraic view of the occurring operations and we find protocols optimizing the tradeoff between the gain in fidelity and the probability of a successful run.
\section{Preliminary remarks}
\label{sec:preliminary}
An orthonormal basis of the two-qubit Hilbert space is given by the maximally entangled Bell states,
\begin{align}
\ket{\Phi^\pm}
&=
\frac{1}{\sqrt{2}}
\left(
\ket{\uparrow \uparrow} \pm \ket{\downarrow \downarrow}
\right),
\\
\ket{\Psi^\pm}
&=
\frac{1}{\sqrt{2}}
\left(
\ket{\uparrow \downarrow} \pm \ket{\downarrow \uparrow}
\right),
\end{align}
where the two spin eigenstates $\ket{\uparrow} \equiv \ket{0}$ (up) and $\ket{\downarrow} \equiv \ket{1}$ (down) define the computational basis. The overlap $F$ of an arbitrary two-qubit quantum state $\rho$ with the Bell state $\ket{\Psi^-}$, i.e.
\begin{equation}
F
\equiv
\bra{\Psi^-} \rho \ket{\Psi^-},
\end{equation}
is referred to as the \textit{fidelity} of the state  $\rho$ in the following. 

Recurrence protocols work on two or more qubit pairs of low fidelity as input that are used to create a single qubit pair with higher fidelity as output. Thereby, only local unitary operations, measurements, and two-way communication of the measurement results via a classical channel can be used. Having initially many copies of the low-fidelity pairs and running the distillation protocol iteratively on the output pairs with higher fidelity, one can achieve fidelities arbitrarily close to $F=1$ and thus, obtain a maximally entangled state.

The original idea of entanglement distillation was introduced in Ref.~\citenum{PhysRevLett.76.722}, and will be referred to as the \textsc{bbpssw} protocol in the following. Initially, the physical setup is such that the two communicating parties, commonly referred to as Alice and Bob, have access to mixed two-qubit states $\rho_\textsc{i}$ of fidelity $F=\bra{\Psi^-} \rho_\textsc{i} \ket{\Psi^-}<1$ that can originate from imperfect sources or noisy quantum channels. To apply the distillation protocol, the state $\rho_\textsc{i}$ first needs to be brought into the Bell-diagonal form \begin{align}
\rho_F
&=
F \ket{\Phi^+} \bra{\Phi^+}
\nonumber \\
&+
\frac{1-F}{3} \Big(
\ket{\Psi^+} \bra{\Psi^+} + \ket{\Psi^-} \bra{\Psi^-} + \ket{\Phi^-} \bra{\Phi^-}
\Big).
\label{eq:rhoF}
\end{align}
This can be achieved for an arbitrary two-qubit state by a so-called \textit{twirl} operation \cite{PhysRevLett.76.722,PhysRevA.54.3824} that retains the component of the rotationally invariant state $\ket{\Psi^-}$, equalizes the components of the other three Bell states, and removes all off-diagonal elements. Thereby Alice and Bob have to implement a random bilateral rotation, i.e.~they choose a random SU(2) rotation and apply it locally to each of the qubits, respectively. As an intermediate result, a so-called Werner state $W_F$ \cite{PhysRevA.40.4277} is created,
\begin{align}
W_F
&=
F \ket{\Psi^-} \bra{\Psi^-}
\nonumber \\
&+
\frac{1-F}{3} \Big(
\ket{\Psi^+} \bra{\Psi^+} + \ket{\Phi^+} \bra{\Phi^+} + \ket{\Phi^-} \bra{\Phi^-}
\Big),
\end{align}
that can be brought into the form in Eq.~(\ref{eq:rhoF}) by performing a unilateral rotation of $\pi$ about the $y$ axis on the Bloch sphere of one of the two qubits, thereby interchanging the $\ket{\Psi^-}$ and $\ket{\Phi^+}$ components.

\section{Asymmetric entanglement distillation with 2 pairs of spins}
\label{sec:1}
%and \cite{RefJ}
\subsection{General interactions between two spins 1/2}
\label{sec:general_distillation}
We start our description with a distillation scheme similar to the original \textsc{bbpssw} protocol \cite{PhysRevLett.76.722}. We replace the symmetric bilateral \textsc{cnot} gate by an asymmetric bilateral operation, in which each local two-qubit operation between qubits $i$ and $j$ is generated from the (an)isotropic exchange interaction
\begin{equation}
\label{eq:Hamiltonian}
H_{ij}(t)
=
\frac{1}{4} J(t)
\left(
\sigma_x^{(i)}\sigma_x^{(j)} + \sigma_y^{(i)}\sigma_y^{(j)}
+ \xi \sigma_z^{(i)}\sigma_z^{(j)}
\right),
\end{equation}
where the $\sigma_\mu^{(i)}$ ($\mu = x,y,z$) are the Pauli matrices describing the $i$th qubit. The parameter $\xi$ quantifies the anisotropy of the interaction, e.g.~for $\xi=1$ the Hamiltonian in Eq.~(\ref{eq:Hamiltonian}) describes isotropic exchange interaction (see Sec.~\ref{subsec:Heisenberg}). The time evolution generated by $H_{ij} (t)$
is\footnote{Here, we set $\hbar = 1$, and time-ordering in Eq.~(\ref{eq:time_evolution}) is not necessary since $[H_{ij} (t), H_{ij} (t')] = 0$ for all $t$ and $t'$.}
\begin{align}
\label{eq:time_evolution}
U_{ij} (\alpha)
&=
e^{-i \int_0^t \mathrm{d}t'\, H_{ij}(t')}
\nonumber \\
&=
\begin{pmatrix}
e^{- i \frac{\alpha \xi}{4}} & 0 & 0 & 0 \\
0 & e^{i \frac{\alpha \xi}{4}} \cos \left( \frac{\alpha}{2} \right) &- i e^{i \frac{\alpha \xi}{4}} \sin \left( \frac{\alpha}{2} \right) & 0 \\
0 & - i e^{i \frac{\alpha\xi}{4}} \sin \left( \frac{\alpha}{2} \right) & e^{i \frac{\alpha\xi}{4}} \cos \left( \frac{\alpha}{2} \right) & 0 \\
0 & 0 & 0 & e^{- i \frac{\alpha\xi}{4}}
\end{pmatrix}.
\end{align}
Here, the matrix representation is in the product basis $\{ \ket{\uparrow \uparrow}, \ket{\uparrow \downarrow},\ket{\downarrow \uparrow},\ket{\downarrow \downarrow}\}$ and we set the initial time to zero. The time evolution is parametrized by the so-called pulse area $\alpha$ defined as
\begin{equation}
\alpha
=
\int \limits_0^t
\mathrm{d}t'\, J(t').
\end{equation}
\begin{figure}
\resizebox{0.5\textwidth}{!}{%
  \includegraphics{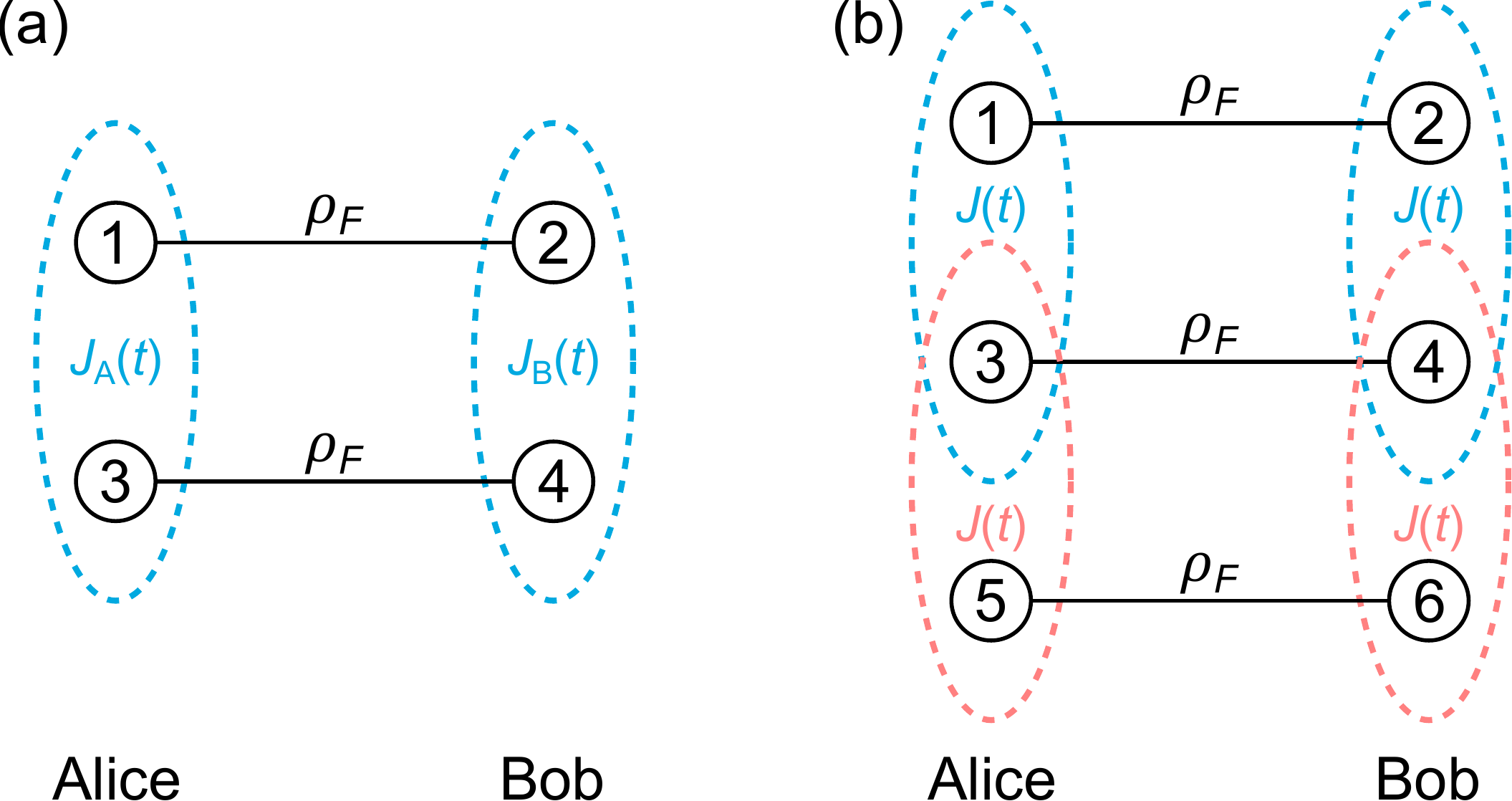}
}
\caption{Entanglement distillation with the exchange interaction. (a) Two input pairs. The parties Alice and Bob share two imperfectly entangled input pairs $\rho_F$, qubit labels are specified. Different exchange pulses generated by the interactions $J_\textsc{a} (t)$ (Alice) and $J_\textsc{b} (t)$ (Bob) implement an asymmetric bilateral two-qubit operation. (b) Three input pairs. Alice and Bob have control on interactions between next local neighbors. Protocols are implemented by iteration of local two-qubit operations.}
\label{fig:setup} 
\end{figure}

For the distillation of partially entangled states that are produced by a source and subsequently transmitted to Alice and Bob, they can use two copies of the state $\rho_F$ [Eq.~(\ref{eq:rhoF})], which can be produced from arbitrary two-qubit states with fidelity $F$ as described in Sec.~\ref{sec:preliminary}. The qubits at Alice' site are labelled 1 and 3, and Bob possesses qubits 2 and 4 (see Fig.~\ref{fig:setup}). The four-qubit state $\rho$ describing such a system is thus given by
\begin{equation}
\rho=\rho_F^{(12)} \otimes \rho_F^{(34)},
\end{equation}
where $\rho_F^{(ij)}$ denotes the state of qubits $i$ and $j$. Afterwards, Alice and Bob each apply an exchange pulse between their respective qubits, which is described by the unitary transformation%
\footnote{We can separate the time evolution of the four-particle system in Eq.~(\ref{P3:eq:time-evolution_total}) into the two-particle propagators $U_{13} (\alpha)$ and $U_{24} (\beta)$ because the Hamiltonians describing each exchange interaction commute, i.e.~$[H_{13}(t),H_{24}(t')]=0$ for all $t$ and $t'$, with $H_{ij} (t)$ given in Eq.~(\ref{eq:Hamiltonian}).}
\begin{equation}
\label{P3:eq:time-evolution_total}
U(\alpha,\beta)
=
U_{13} (\alpha) \otimes U_{24} (\beta).
\end{equation}
If we denote the exchange couplings in Alice's and Bob's spin register as $J_\textsc{a} (t)$ and  $J_\textsc{b} (t)$, then the pulse areas $\alpha$ and $\beta$ are given by 
\begin{align}
\alpha 
&=
\int_0^{t_\textsc{a}} \mathrm{d} t'\, J_\textsc{a} (t'),
\\
\beta
&=
\int_0^{t_\textsc{b}} \mathrm{d} t'\, J_\textsc{b} (t'),
\end{align}
where $t_\textsc{a/b}$ are the respective pulse lengths. The crucial difference to other entanglement distillation protocols \cite{PhysRevA.78.062313,PhysRevA.78.022312,PhysRevA.84.042303,PhysRevA.86.052312,PhysRevLett.94.236803,Pan:2001} is that Alice and Bob are allowed to choose different pulse areas and thus, apply different bilateral two-qubit operations. It is exactly this asymmetric bilateral operation that makes entanglement distillation via exchange interaction of the form in Eq.~(\ref{eq:Hamiltonian}) feasible \textit{at all} if only two input pairs are used. The exchange pulses transform the four-qubit state $\rho$ as
\begin{equation}
\rho 
\mapsto
U(\alpha,\beta) \rho U(\alpha,\beta)^\dagger.
\end{equation}
After this unitary transformation, the two parties continue in the same way as in the original \textsc{bbpssw} protocol. Although we do not use any conditional quantum operations here, we still denote qubits 1 and 2 as control qubits, and qubits 3 and 4 as target qubits. Alice and Bob measure the target qubits in the computational basis $\{\ket{\uparrow},\ket{\downarrow}\}$ and compare the measurement results afterwards using classical two-way communication. If Alice and Bob obtain equal measurement results, i.e. either both spins are pointing up or both are pointing down, they will keep the control qubits. Otherwise, the state is discarded. 
In case that the control qubits are kept, another unilateral rotation of $\pi$ about the $y$ axis on the Bloch sphere is applied to interchange again the $\ket{\Psi^-}$ and $\ket{\Phi^+}$ components.
As we derive below, in case of keeping the control pair, the fidelity of precisely this state can become larger than the initial fidelity $F$ through the above transformation and measurement, depending on the applied exchange pulses $\alpha$ and $\beta$.

If we denote the postselected state of the control qubits by $\rho'$, then the output fidelity $F_\textrm{out}(F,\alpha,\beta) =\bra{\Psi^-} \rho' \ket{\Psi^-}$ is found to be
\begin{equation}
\label{eq:fidelity}
F_\textrm{out} (F, \alpha, \beta)
=
\frac{\nu(F, \alpha, \beta)}{\delta(F, \alpha, \beta)},
\end{equation}
with
\begin{align}
&\nu(F, \alpha, \beta)
=
3 (4F-1) \cos(\alpha)\cos(\beta) 
\nonumber \\
&+ 4 (8F^2 + 2F -1) \cos \left( \frac{\alpha + \beta}{2} \xi \right) \cos \left( \frac{\alpha + \beta}{2} \right)
\nonumber \\
&-(4F-1)^2 \sin(\alpha) \sin(\beta)
+4F (4F+1) + 7,
\end{align}
and
\begin{align}
\delta(F, \alpha, \beta)
&=
6(4F-1) \cos (\alpha) \cos  (\beta) 
\nonumber \\
&- 2(4F-1)^2 \sin (\alpha) \sin (\beta) + 6(4F+5)
\end{align}
A detailed analysis of the fidelity $F_\textrm{out} (F, \alpha, \beta)$ may be found in Ref.~\citenum{PhysRevA.90.022320} for the isotropic case $\xi = 1$. An interesting property is found when Alice and Bob apply mutually inverse operations, i.e.~$\alpha = -\beta$. In this case, the fidelity $F_\textrm{out} (F, \alpha, -\alpha)$ becomes independent of the anisotropy parameter $\xi$,
\begin{align}
&F_\textrm{out}(F, \alpha, - \alpha)
=
\frac{1}{2} 
\nonumber \\
&+\frac{3 - 12 F^2}{(F - 1)(4F - 1) \cos(2 \alpha) - F(4F + 7) - 7}
\label{eq:fidelity_inverse_operations}
\end{align}
Since in this case, a repulsive fixed point $F_\textrm{min}=1/2$ and an attractive fixed point $F_\textrm{max}=1$ of the map $F_\textrm{out} (F, \alpha, -\alpha)$ are found, it allows the distillation of maximally entangled states. In the range $1/2<F<1$, the maximum of $F_\textrm{out}(F, \alpha, - \alpha)$ is obtained for $\alpha = \pi/2$.
\begin{figure}
\resizebox{0.5\textwidth}{!}{%
\includegraphics{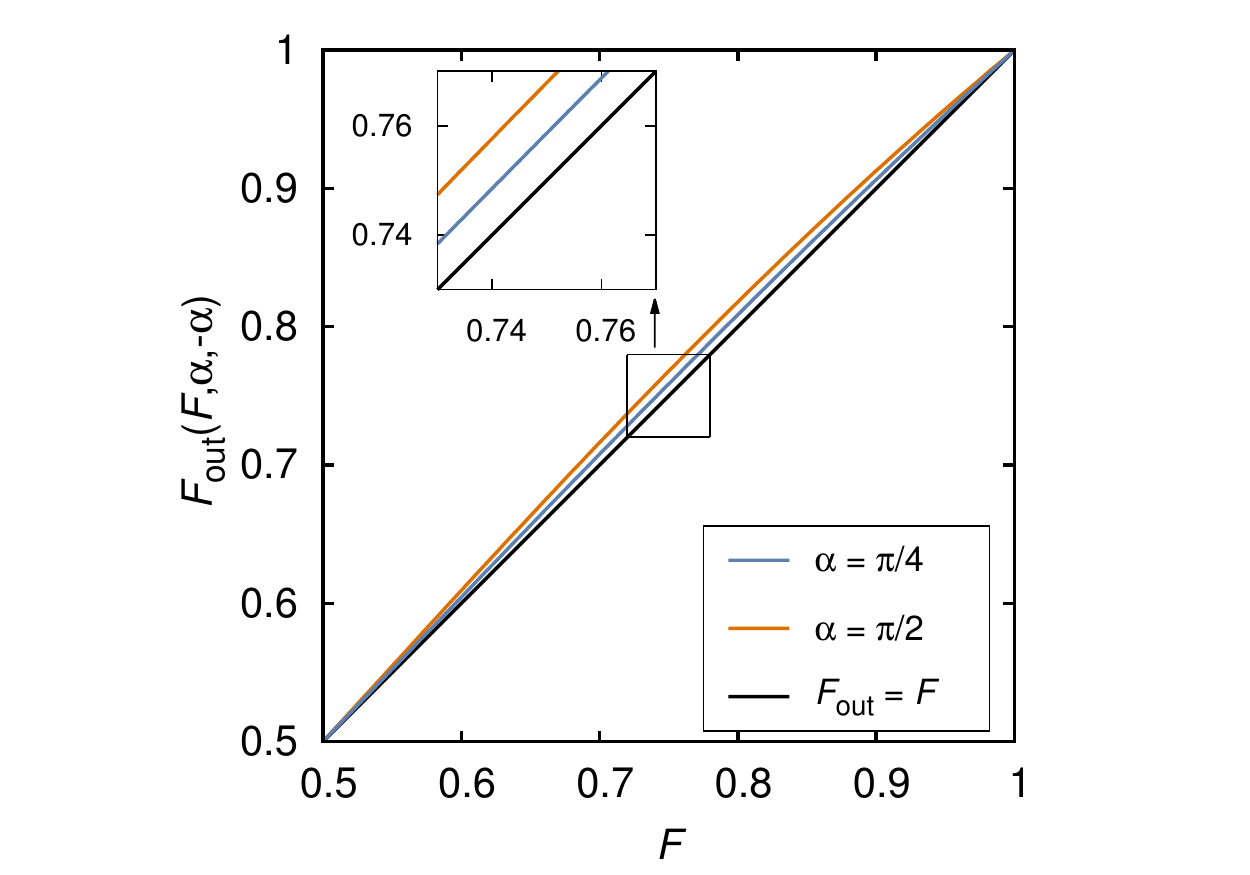}
}
\caption{Increased fidelity $F_\textrm{out}(F,\alpha,-\alpha)$ after a single distillation step as a function of $F$, here shown for pulse areas $\alpha = \pi/4$ and $\alpha = \pi/2$.}
\label{fig:fidelity} 
\end{figure}
\subsection{Heisenberg exchange interaction}
\label{subsec:Heisenberg}
Isotropic exchange interaction ($\xi = 1$) is described by a Heisenberg Hamiltonian, i.e.
\begin{equation}
H_{ij}^{\textsc{h}} (t)
=
\frac{1}{4} J(t) \bm{\sigma}_i \cdot \bm{\sigma}_j,
\label{eq:Heisenberg}
\end{equation}
where $\bm{\sigma}$ denotes the vector of Pauli matrices. Typically, electron spins in gate-defined quantum dots are coupled by such an interaction that can be used to implement universal quantum computation \cite{PhysRevA.57.120,PhysRevB.59.2070,Petta30092005,RevModPhys.79.1217,doi:10.1146/annurev-conmatphys-030212-184248}. This case has been studied in Ref.~\citenum{PhysRevA.90.022320}, and it was found that the highest gain in fidelity is given for pulse areas $\alpha = \pi/2 = - \beta$,
\begin{align}
F_\textrm{out,opt}(F)
&\equiv
F_\textrm{out} \left(F, \frac{\pi}{2}+ 2\pi n, -\frac{\pi}{2}+ 2\pi m  \right)
\nonumber \\
&
= \frac{16 F^2 + F + 1}{8 F^2 + 2 F + 8},
\label{eq:twoopt}
\end{align}
where the integers $n$ and $m$ can be chosen independently by Alice and Bob. In the optimal case, Alice applies the so-called $\sqrt{\textsc{swap}}$ gate,
\begin{equation}
U_{\sqrt{\textsc{swap}}}
=
\begin{pmatrix}
1 & 0 & 0 & 0 \\
0 & (1-i)/2 & (1+i)/2 & 0 \\
0 & (1+i)/2 & (1-i)/2 & 0 \\
0 & 0 & 0 & 1
\end{pmatrix},
\end{equation}
and Bob the inverse $\!\sqrt{\textsc{swap}}$ gate, $\!\sqrt{\textsc{swap}}^{-1}$.\footnote{The square of  $\!\sqrt{\textsc{swap}}^{-1}$ is also the \textsc{swap} operation and it can be understood as another root of \textsc{swap}.}
\subsection{XY interaction}
The Hamiltonian describing XY-type interaction is obtained for $\xi = 0$ and thus given by
\begin{equation}
H_{ij}^{\textsc{xy}} (t)
=
\frac{1}{4} J(t)
\left(
\sigma_x^{(i)} \sigma_x^{(j)} + \sigma_y^{(i)} \sigma_y^{(j)}
\right),
\end{equation}
i.e.~only the $x$ and $y$ components of the spins are coupled. This kind of interaction appears, e.g., in all-optical cavity-coupled QD electron spins \cite{PhysRevLett.83.4204} or superconducting qubits \cite{RevModPhys.73.357}. For a pulse area $\alpha = - \pi$, the Hamiltonian $H_{ij}^{\textsc{xy}} (t)$ generates e.g.~the so-called i\textsc{swap} gate,
\begin{equation}
U_{\textrm{i}\textsc{swap}}
=
\begin{pmatrix}
1 & 0 & 0 & 0 \\
0 & 0 & i & 0 \\
0 & i & 0 & 0 \\
0 & 0 & 0 & 1 \\
\end{pmatrix}.
\end{equation}
For distillation, Alice and Bob follow the scheme described in Sec.~\ref{sec:general_distillation}, i.e. they start with two qubit pairs $\rho = \rho_F \otimes \rho _F$ and apply the XY interaction with pulse areas $\alpha$ and $\beta$ to their respective qubit pairs. After a subsequent measurement of the target qubits, Alice and Bob keep the control pair if they obtain equal measurement results. The fidelity $F_{\textrm{out},\textsc{xy}}(F, \alpha, \beta)$ of the source state can be increased depending on the pulse areas $\alpha$ and $\beta$, and a formula for $F_{\textrm{out},\textsc{xy}}(F, \alpha, \beta)$ can be found in Ref.~\citenum{PhysRevA.90.022320}.
As discussed before, In the case $\alpha = - \beta$, i.e.~when both parties apply mutually inverse operations, the result coincides with Eq.~(\ref{eq:fidelity_inverse_operations}), and the gain in fidelity is thus maximal for $\alpha = \pi / 2$. In the optimal case here, the different qubit interactions correspond to gates whose double application result in the i\textsc{swap} gate.
\subsection{Dipole-dipole interaction}
The dipole-dipole coupling between two magnetic moments $\bm{\mu}_i = \gamma \textbf{S}_i$ and $\bm{\mu}_j = \gamma \textbf{S}_j$, separated by a distance $r$, is described by the Hamiltonian \cite{Abragam1961}
\begin{equation}
\label{P3:eq:Hamiltonian_dipole}
H_\textrm{dd}
=
\frac{\mu_0 \gamma^2}{4 \pi r^3}
\left(
\textbf{S}_i \cdot \textbf{S}_j - 3
\left(
\textbf{S}_i \cdot \textbf{e}_r
\right)
(
\textbf{S}_j \cdot \textbf{e}_r
)
\right).
\end{equation}
Here $\textbf{e}_r$ is a unit vector pointing along the connecting line between the two identical magnetic moments with gyromagnetic ratio $\gamma$ and $\mu_0$ is the vacuum permeability. For example, the electron spins of two nitrogen-vacancy centers in diamond that are close enough to each other can be coupled via the interaction of the associated magnetic moments and entangled in this way \cite{Neumann:2010}. Without loss of generality, we can assume the connecting line to define the $z$ axis and thus, obtain
\begin{equation}
H_\textrm{dd}
=
\frac{\mu_0 \gamma^2}{16 \pi r^3}
\left(
\sigma_x^{(i)} \sigma_x^{(j)} + \sigma_y^{(i)} \sigma_y^{(j)} - 2\sigma_z^{(i)} \sigma_z^{(j)}
\right),
\end{equation}
where we assume spin-1/2 systems that are magnetically coupled. The Hamiltonian $H_\textrm{dd}$ is thus obtained for anisotropy parameter of $\xi = - 2$. The strength of the interaction could in principle be varied by changing the distance $r$ between the qubits, which might not be a trivial task. However, as proof of principle of our developed concept and to demonstrate that it works for a variety of Hamiltonians, we apply the asymmetric distillation scheme developed above as well to qubits coupled via $H_\textrm{dd}$. We define the pulse area as $\alpha = \int_0^t \mathrm{d} t'\, \mu_0 \gamma^2/(16 \pi r(t')^3)$ and assume a time-dependent distance $r(t)$. The fidelity $F_\textrm{out,dd}(F,\alpha,\beta)$ after one distillation round with initial fidelity $F$ is calculated to be
\begin{equation}
F_\textrm{out,dd}(F,\alpha,\beta)
=
\frac{\nu_\textrm{dd}(F,\alpha,\beta)}
{\delta(F,\alpha,\beta)} ,
\end{equation}
and the numerator $\nu_\textrm{dd}(F,\alpha,\beta)$ is given by the expression
\begin{align}
\nu_\textrm{dd}(F,\alpha,\beta)
&=
(2 F(4F+1) - 1)
[
2 \cos(2(\alpha+\beta))
\nonumber \\
&+
\cos(4(\alpha+\beta))
+
2 \cos(6(\alpha+\beta))
]
 \nonumber \\
 &
- 2 (F-1)(4F-1)\cos(4(\alpha-\beta))
\nonumber \\
&+4F(4F + 1) + 7.
\end{align}
Upon detailed inspection, one finds
\begin{equation}
F_\textrm{out,dd}\left(F,\frac{\alpha}{4},-\frac{\alpha}{4}\right)
=
F_\textrm{out}(F,\alpha,-\alpha),
\end{equation}
and therefore the discussion of Sec.~\ref{subsec:Heisenberg} also applies for entanglement distillation in case of qubits coupled via magnetic dipole-dipole interaction, with the optimal distillation achieved for a pulse area of $\alpha = \pi / 8$.
\section{Symmetric entanglement distillation with 3 or more pairs of spins}
\subsection{Extension to three qubit pairs }
\label{sec:general_3bit}
\begin{figure}
\centering
\includegraphics[width=0.98\linewidth]{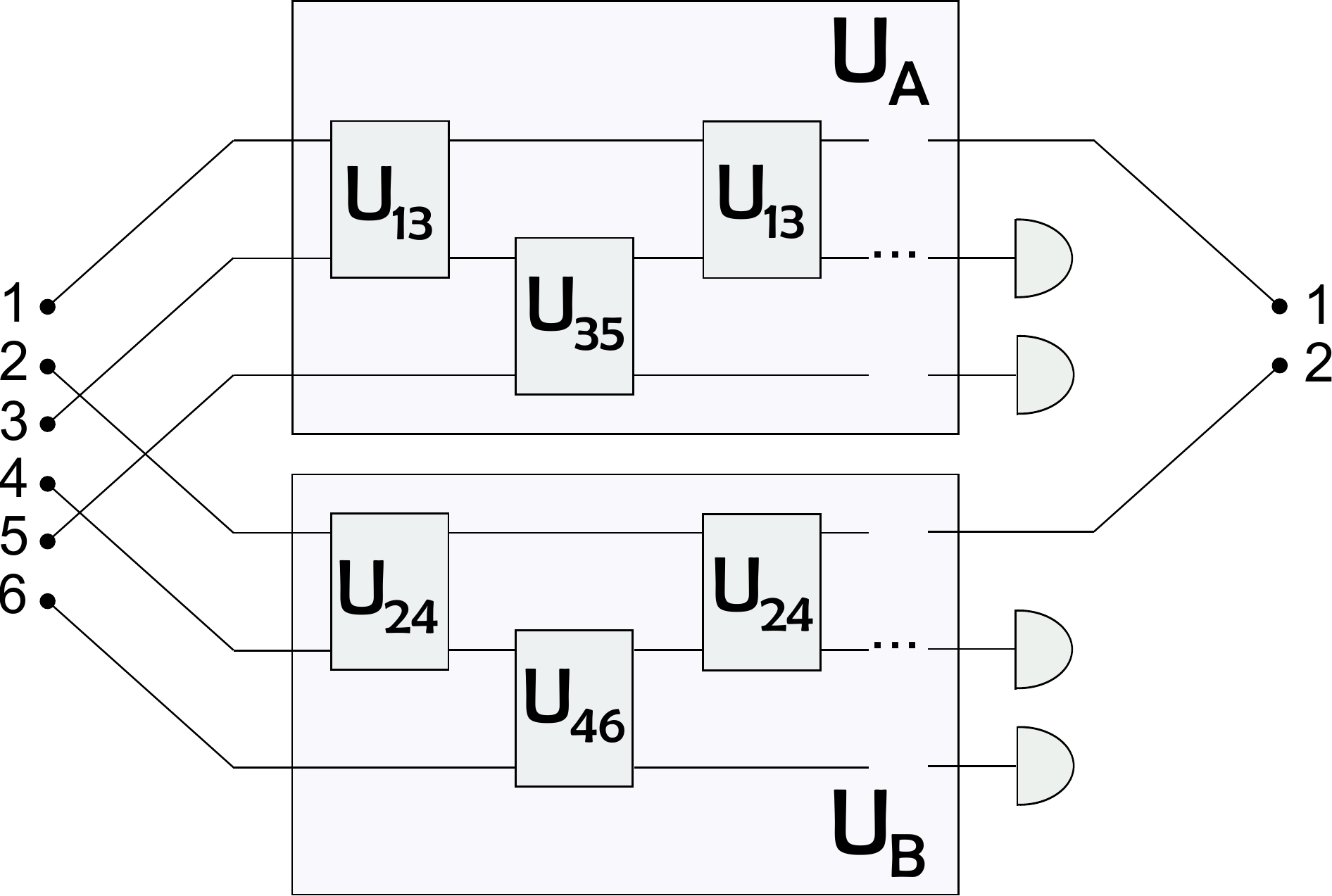}
\caption{
Circuit of protocols implemented by Alice and Bob via iterated applications of controlled exchange interactions on the qubits $(1,3,5)$ and $(2,4,6)$ followed by a measurement on the qubits $(3,4,5,6)$.
}
\label{fig:protocol}
\end{figure}

In this section we will extend the above setting to a scenario where Alice and Bob have access to three bipartite qubit pairs in a global state $\rho_F^{\otimes 3}$ and local control on isotropic exchange interactions between next nearest neighbors, see Fig.~\ref{fig:setup}. We number the qubits  from $1$ to $6$, where Alice has access to odd numbers and can control exchange interactions between the qubit pairs $(1,3)$ and $(3,5)$. Analogously, Bob has access to even numbered qubits and controls interactions between the pairs $(2,4)$ and $(4,6)$.\\
As in Sec.~\ref{sec:general_distillation}, we will consider protocols where both parties first apply controlled sequences of local exchange interactions, resulting in overall unitary operations $U_A$ and $U_B$\footnote{For clarity of notation: the unitaries $U_A$ and $U_B$ are assumed to be represented as matrices on $(\mathbb{C}^2)^{\otimes 6}$ with $U_A$ acting as identity on Bob's qubits and $U_B$ as identity on Alice's. }. Then they apply a filter based on one round of classical communication. This filter is implemented by measuring each of the qubits $3-6$ in the computational basis and keeping the state of the qubit pair $(1,2)$ whenever the measurements on the qubit pairs $(3,4)$ and $(5,6)$ coincide, see Fig.~\ref{fig:protocol}. This is described by the projection
\begin{align}
  \Ppara&=\bigl(\ketbra{0_30_4}+ \ketbra{1_31_4}\bigr)\nonumber\\
        &\strut\qquad\otimes\bigl(\ketbra{0_50_6}+ \ketbra{1_51_6}\bigr),
\end{align}
and the output state $\rho\out$ with fidelity $F\out$ relative to the maximally entangled target state $\ketbra{\Phi^+}$ is obtained with success probability $\Psucc$, given by
\begin{align}
\Psucc&:=\operatorname{tr}(\mathbb{I}_{12}\otimes \Ppara\, U_A U_B \rho_F^{\otimes 3} U_B^\dagger U_A^\dagger) \\
\rho\out&:=\frac{1}{\Psucc}\operatorname{tr}_{3,4,5,6}
           (\mathbb{I}_{12}\otimes \Ppara\, U_A U_B \rho_F^{\otimes 3} U_B^\dagger U_A^\dagger) \\
F_\textrm{out}&:=\frac{1}{\Psucc}\operatorname{tr}\left(\ketbra{\Phi^+}\otimes P_\| \, U_A U_B \rho_F^{\otimes 3}  U_B^\dagger U_A^\dagger \right) .
\end{align}

\subsection{The reachable set of unitaries}
\label{sec:reachable_set}
At first we will have to characterize the reachable set of unitaries, $U_A$ and $U_B$, which could be implemented by sequences of Alice's and Bob's basic operations. This characterization can be done separately for the two parties and we will only consider Alice's side explicitly.

Alice can implement the basic operations, see \eqref{eq:Heisenberg},
\begin{align}
 e^{-i\int_0^t \mathrm{d}t'\, H_{13}^\text{H}(t')}\quad \text{and}\quad
 e^{-i\int_0^t \mathrm{d}t'\, H_{35}^\text{H}(t')},
\end{align}
by switching on and off an isotropic exchange interaction for a specific time $t$. Up to an irrelevant global phase, these operations can be expressed \cite{PhysRevA.90.022320}, as
\begin{align}
U_{13}(\xi)&:=e^{i \xi \mathbb{F}_{13}} \nonumber\\
U_{35}(\chi)&:=e^{i \chi \mathbb{F}_{35}},
\label{eq:generators}
\end{align}
where $\mathbb{F}_{ij}$ denotes the flip operation, i.e. it permutes the $i$th and $j$th  tensor factor in $(\mathbb{C}^2)^{\otimes 6}$.
If Alice iterates the operations \eqref{eq:generators} with time steps $\{\xi_i\}$ and $\{\chi_i\}$ all unitaries she can implement are of a form
\begin{align}
U_A=\prod_{i} U_{13}(\xi_i)U_{35}(\chi_i)=
\prod_{i} e^{i \xi_i \mathbb{F}_{13}}e^{i \chi_i \mathbb{F}_{35}}.
\label{eq:product}
\end{align}

The idea for simplifying long products of such operators, with judiciously chosen parameters $\xi$ and $\chi$ is to utilize the commutation relations of the flip operators. Indeed, if the exponentials are expanded, each factor will be a product of the operators $\mathbb{F}_{13}$ and $\mathbb{F}_{35}$, and these can all be evaluated to some permutation operator of the three sites $(1,3,5)$, i.e., one of the operators $\{\mathbb{I},\mathbb{F}_{13},\mathbb{F}_{35},\mathbb{F}_{15},\mathbb{Z}_{135},\mathbb{Z}_{153}\}$, where $\mathbb{Z}_{ijk}$ denotes the (anti-)cyclic permutation. These operators span a finite dimensional algebra $\mathcal{A}$, for which a convenient basis \cite{PhysRevA.63.042111} is
\begin{align}
A_+&=\frac 1 6 \left(
 \mathbb{I}+\mathbb{F}_{13}+\mathbb{F}_{35}+\mathbb{F}_{15}+\mathbb{Z}_{135}+\mathbb{Z}_{153}
 \right)
 \nonumber\\
A_-&=\frac 1 6 \left(
 \mathbb{I}-\mathbb{F}_{13}-\mathbb{F}_{35}-\mathbb{F}_{15}+\mathbb{Z}_{135}+\mathbb{Z}_{153}
 \right)
 \nonumber\\
A_0&=\frac{1}{3}\left(2\mathbb{I}-\mathbb{Z}_{135}-\mathbb{Z}_{153}\right)
\nonumber\\
A_1&=\frac{1}{3}\left(2\mathbb{F}_{35}-\mathbb{F}_{13}-\mathbb{F}_{15}\right)
\nonumber\\
A_2&=\frac{1}{\sqrt 3}\left(\mathbb{F}_{13}-\mathbb{F}_{15}\right)
\nonumber\\
A_3&=\frac{i}{\sqrt 3}\left(\mathbb{Z}_{135}-\mathbb{Z}_{153}\right).
\label{eq:sureweilAlice}
\end{align}
Here $A_+$, $A_-$ and $A_0$ are three orthogonal projectors summing up to $\mathbb{I}$. They correspond to different irreducible representations of the permutation group acting on three qubits. $A_+$ and $A_-$ are the trivial and the alternating representation, which act as projectors on the symmetric and antisymmetric subspace.
$A_0$ corresponds to a two dimensional representation on which the matrices $A_1$, $A_2$ and $A_3$ act as Pauli matrices, i.e. $[A_l,A_m]=2i\epsilon_{lmn}A_n$ and $A_1^2=A_2^2=A_3^2=A_0\bot A_+$.

Now any product of a form as in \eqref{eq:product} can be computed
in the basis \eqref{eq:sureweilAlice} yielding a unitary that is in the algebra $\mathcal{A}$. As there is no fully antisymmetric state of three qubits we do not further have to take into account $A_-$. Hence $A_0$ acts like $\mathbb{I}-A_+$ such that, up to an irrelevant global phase, \eqref{eq:product} can always be written as
\begin{align}
U_A=e^{i(\alpha A_+ + \bm{a}\cdot\bm{A})},
\label{eq:UA}
\end{align}
with parameters $\alpha\in(0,2\pi)$ and the vectors $\bm{a} \in \mathbb{R}^3$ such as $\bm{A}=(A_1,A_2,A_3)$.

Likewise Bob's unitaries are described by parameters $\beta\in(0,2\pi)$ and $\bm{b} \in \mathbb{R}^3$
as
\begin{align}
U_B=e^{i(\beta B_+ + \bm{b}\cdot\bm{B})},
\label{eq:UB}
\end{align}
with $B_+$ and $\bm{B}$ are defined on the qubits $(2,4,6)$ in the same manner as for Alice.

In our case also the converse holds: Every unitary of the form \eqref{eq:UA} can be obtained as a product as in \eqref{eq:product}. The basic criterion for this is that the operators $\mathbb I$, $\mathbb{F}_{13}$ and $\mathbb{F}_{35}$ and their iterated commutators span the whole algebra $\mathcal A$ \cite[theorem 2.3]{GAMM:GAMM200890003}.  Finding an explicit and efficient decomposition is in general a complicated task which is the subject of control theory. A good introduction to this interesting topic can be found in \cite{Brockett:1973,GAMM:GAMM200890003}.

\subsection{Pretty good protocols}
\begin{figure}[h]
\includegraphics[width=0.95\linewidth]{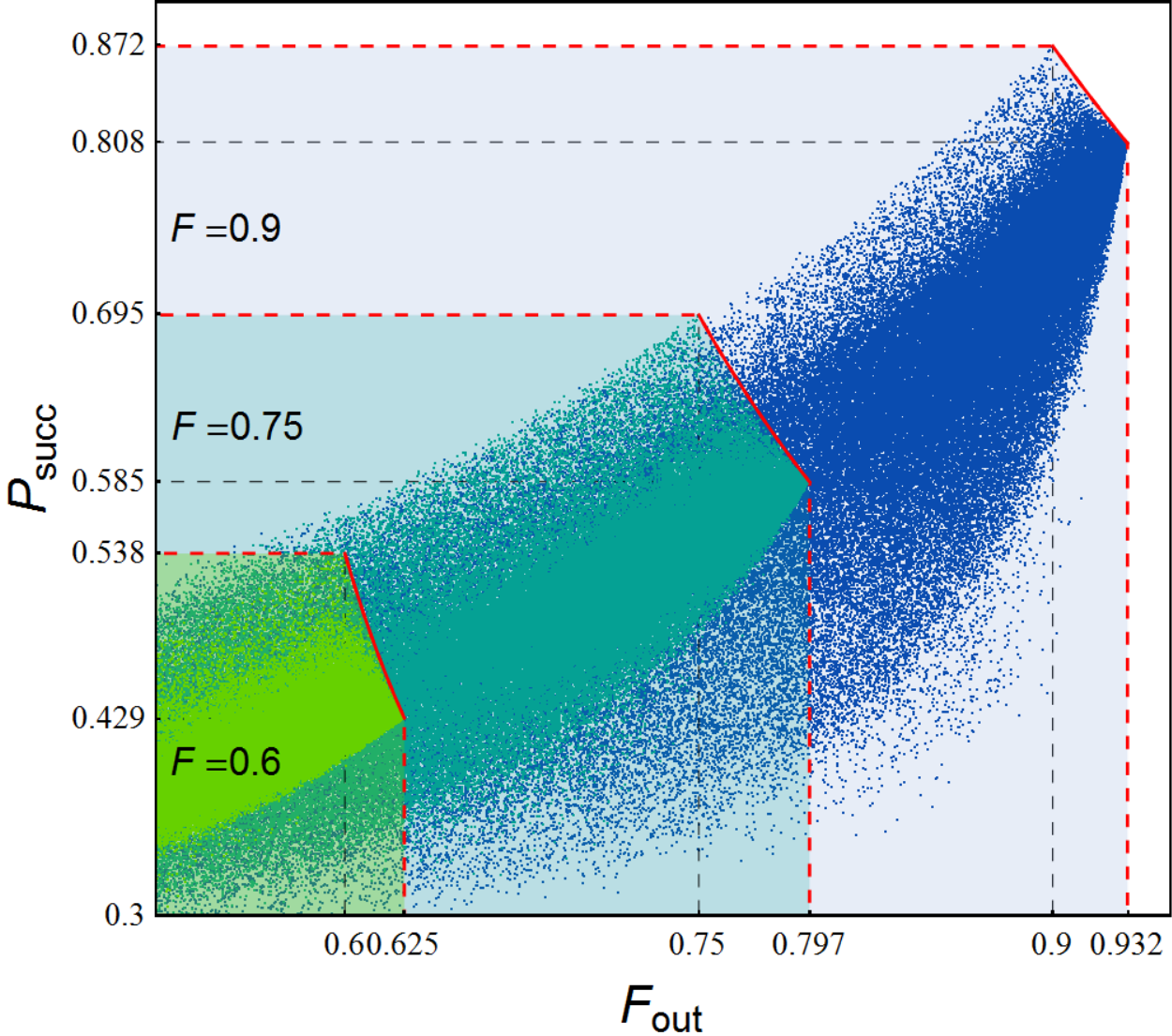}
\caption{
Random sample of $F_\textrm{out}$ and $\Psucc$ for achievable protocols with input fidelities $F\in(0.6,0.75,0.9)$. The respective Pareto fronts are marked in red. The Pareto dominated region is bounded by dashed red lines.
The sample size is $2\cdot10^5$ for each case.
}
\label{fig:sample}
\end{figure}

%-what is a good protocol
We can get a qualitative overview over the attainable characteristics of possible protocols by random sampling of $U_A$ and $U_B$, i.e., by choosing $(\alpha,\bm a,\beta,\bm b)$ in \eqref{eq:UA} at random.  Fig.~\ref{fig:sample} shows such a sample of attainable values of $F\out$ and $\Psucc$ for different fixed input fidelities $F$.
Good protocols in this set are those with a favorable trade-off between $F\out$ and $\Psucc$.
This can be made precise by the notion of {\it Pareto efficiency} \cite{chinchuluun2007}: We say that one protocol {\it dominates} another whenever it attains higher fidelity and a higher success probability, and at least one of these parameters is even strictly higher. A protocol which can not be dominated by any other is said to be {\it Pareto efficient}, and  the corresponding set of pairs $(F_\textrm{out},\Psucc)$ attained by Pareto efficient protocols is called the {\it Pareto front}. By definition, the front is a tradeoff curve, along which higher fidelity means lower success probability and conversely.

\begin{figure}
\centering
\includegraphics[width=0.9\linewidth]{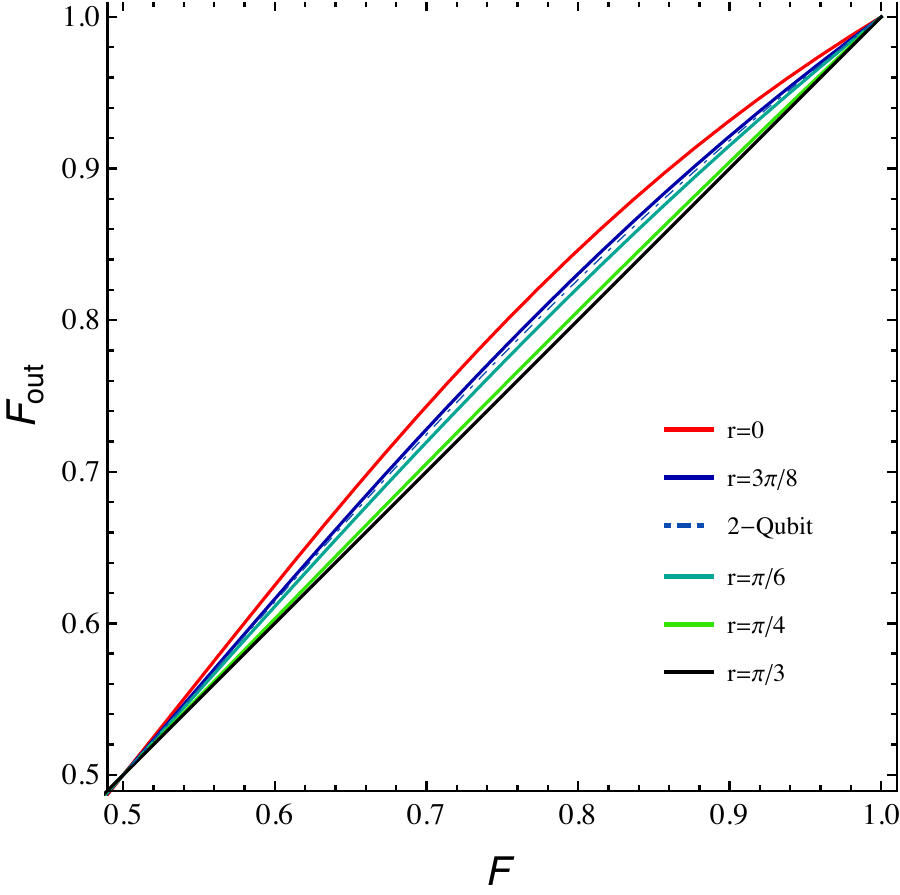}\\
\vspace{0.5cm}
\includegraphics[width=0.9\linewidth]{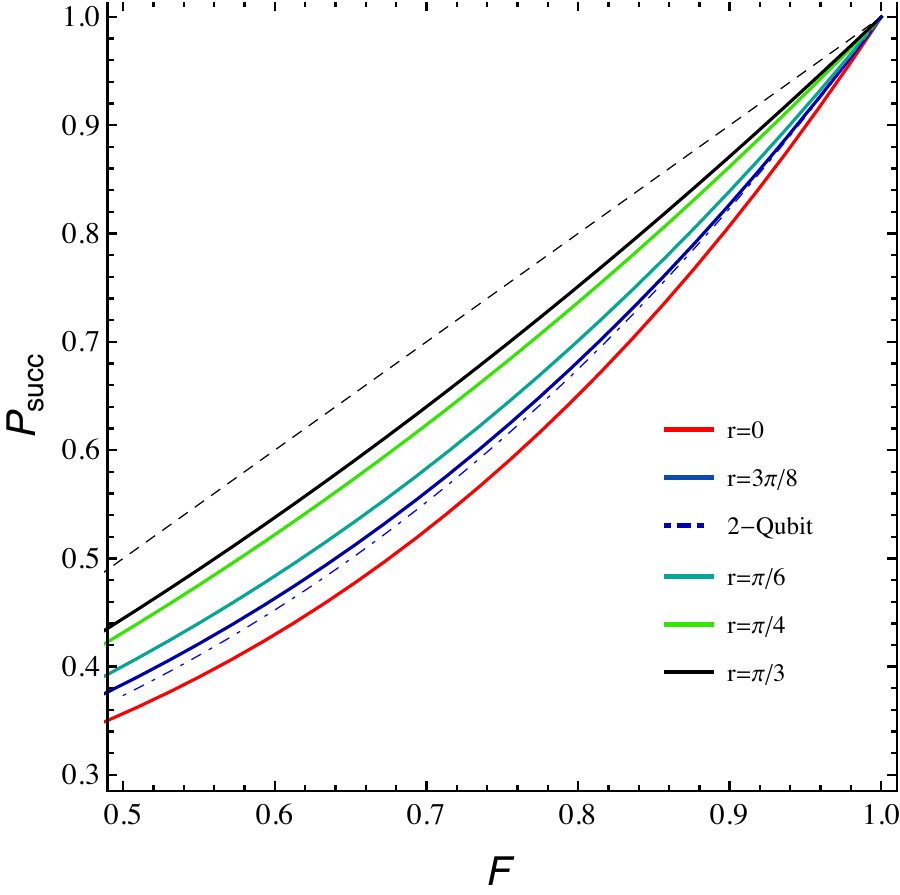}
\caption{
Output fidelity and success probability of the Pareto efficient protocols from \eqref{eq:effiprot}(solid lines) .
Output fidelity and success probability of the optimal two quit protocol \eqref{eq:twoopt} iterated on three qubit pairs (dotdashed line).
}
\label{fig:Fid_Prob}
\end{figure}

By numerical optimization we identify a family of Pareto efficient protocols, as those with parameters
\begin{align}
\alpha=\beta=\pi \quad \bm{a}=\bm{b}=(0,0,r),
\end{align}
with $r\in(0,\pi/3)$. Remarkably, these protocols are, as in the case of two qubit pairs, independent of the input fidelity $F$.
The output fidelity and the success probability describing the Pareto front is shown in Fig.~\ref{fig:Fid_Prob} and can be computed as
\begin{align}
F_\textrm{out}&=
\frac
{16 \left(4 F^2-5 F+1\right) \cos (3 r)+226 F^2+F+16}
{32 \left(4 F^2-5 F+1\right) \cos (3 r)+128 F^2+2 F+113}
\\
\Psucc&=\frac{(2 F+1)}{729}\nonumber\\
&\times\left(32 \left(4 F^2-5 F+1\right)\cos(3r)+128 F^2+2 F+113\right).\nonumber\\
\quad
\label{eq:effiprot}
\end{align}
For $r=\pi/3$ the highest success probability and the lowest output fidelity is attained. In this case the efficiency equals the case in which Alice and Bob apply no interaction at all.
In contrast, for $r=0$ the highest output fidelity and the lowest success probability is attained and the maximal achievable fidelity can be computed as
\begin{align}
F_\textrm{max}=\frac{290 F^2-79 F+32}{256 F^2-158 F+145}.
\end{align}

As a last point we can compare the Pareto efficient protocols from \eqref{eq:effiprot}, with an iteration of the optimal two qubit pair protocol \eqref{eq:twoopt} acting on three qubit pairs. This is shown in Fig.~\eqref{fig:Fid_Prob}. We can see that for every input fidelity $F>1/2$ there is a Pareto efficient protocol from the family  \eqref{eq:effiprot} which attains an equal or bigger fidelity gain with a higher success probability. Moreover also a higher fidelity gain is possible when a lower success probability is accepted. Nevertheless one always has to keep in mind that, in the above setting, perfectly controlled sequences of interactions are assumed. Hence this might harder to realize in an experiment, than an iterated two qubit protocol, which can be implemented by only two steps of controlled interactions.

\section{Conclusions and Outlook}
We presented entanglement distillation protocols based on the exchange interaction using either two or three input pairs. In the case of two input pairs, we analyzed a protocol based on (an)isotropic exchange and found that entanglement distillation is possible for various interaction types, namely Heisenberg exchange, XY interaction and dipole-dipole interaction. If Alice and Bob apply mutually inverse operations, it turns out that the output fidelity becomes independent of the anisotropy parameter $\xi$. Further studies could investigate more general spin-spin interactions of the form $\textbf{S}_i^T \cdot \overleftrightarrow{\textbf{J}} \cdot \textbf{S}_j$, with some non-diagonal coupling tensor $\overleftrightarrow{\textbf{J}}$. An example of such an interaction is the so-called Dzyaloshinskii-Moriya interaction \cite{Dzyaloshinsky1958241,PhysRev.120.91}, which arises from spin-orbit coupling.

The above results on three input pairs directly suggest a scheme for finding protocols acting on $n$-qubit pairs by locally controlled next nearest neighbor exchange interactions. The operations $\exp(i\alpha A_+)$ and $\exp(i\beta B_+)$ can indeed be generalized to arbitrary numbers of qubit pairs by choosing $A_+$ and $B_+$ as projectors on the symmetric $n$-particle subspaces. However, one has to consider that with an increasing number of qubit pairs the probability of a joint coincidence of measurements on $n-1$ qubit pairs decreases exponentially. Hence a more detailed investigation is needed to decide whether distillation via exchange interaction can be used to produce, with a positive rate,  almost maximally entangled pairs from a source of sufficiently highly entangled pairs.

\begin{acknowledgement}
A.~A.~and G.~B.~acknowledge funding from the BMBF under the program Q.com-HL and from the DFG within SFB 767. R.~S.~and R.~F.~W.~acknowledge funding from the BMBF under the program Q.com-Q, R.~F.~W.~additionally acknowledges the ERC grand DQSIM, and L.~D.~is funded from the DFG within RTG 1991.
\end{acknowledgement}

% \bibliographystyle{thesis_style.bst}
% \bibliography{../../../Bibliography/bibliography-all}

\begin{thebibliography}{10}

\bibitem{Kimble:2008}
H.~J. Kimble, Nature \textbf{453}, 1023 (2008).

\bibitem{PhysRevLett.67.661}
A.~K. Ekert, Phys. Rev. Lett. \textbf{67}, 661 (1991).

\bibitem{PhysRevLett.81.5932}
H.-J. Briegel, W.~D\"ur, J.~I. Cirac, and P.~Zoller, Phys. Rev. Lett.
  \textbf{81}, 5932 (1998).

\bibitem{PhysRevA.59.169}
W.~D\"ur, H.-J. Briegel, J.~I. Cirac, and P.~Zoller, Phys. Rev. A \textbf{59},
  169 (1999).

\bibitem{Simon:2010}
C.~Simon, M.~Afzelius, J.~Appel, A.~Boyer de~la Giroday, S.~J. Dewhurst,
  N.~Gisin, C.~Y. Hu, F.~Jelezko, S.~Kr\"oll, J.~H. M\"uller, J.~Nunn, E.~S.
  Polzik, J.~G. Rarity, H.~De~Riedmatten, W.~Rosenfeld, A.~J. Shields,
  N.~Sk\"old, R.~M. Stevenson, R.~Thew, I.~A. Walmsley, M.~C. Weber,
  H.~Weinfurter, J.~Wrachtrup, and R.~J. Young, Eur. Phys. J. D \textbf{58}, 1
  (2010).

\bibitem{doi:10.1146/annurev-conmatphys-030212-184248}
C.~Kloeffel and D.~Loss, Annu. Rev. Condens. Matter Phys. \textbf{4}, 51
  (2013).

\bibitem{doi:10.1146/annurev-conmatphys-030212-184238}
V.~Dobrovitski, G.~Fuchs, A.~Falk, C.~Santori, and D.~Awschalom, Annu. Rev.
  Condens. Matter Phys. \textbf{4}, 23 (2013).

\bibitem{PhysRevLett.76.722}
C.~H. Bennett, G.~Brassard, S.~Popescu, B.~Schumacher, J.~A. Smolin, and W.~K.
  Wootters, Phys. Rev. Lett. \textbf{76}, 722 (1996).

\bibitem{PhysRevLett.77.2818}
D.~Deutsch, A.~Ekert, R.~Jozsa, C.~Macchiavello, S.~Popescu, and A.~Sanpera,
  Phys. Rev. Lett. \textbf{77}, 2818 (1996).

\bibitem{PhysRevA.57.120}
D.~Loss and D.~P. DiVincenzo, Phys. Rev. A \textbf{57}, 120 (1998).

\bibitem{Nowack30112007}
K.~C. Nowack, F.~H.~L. Koppens, Y.~V. Nazarov, and L.~M.~K. Vandersypen,
  Science \textbf{318}, 1430 (2007).

\bibitem{Petta30092005}
J.~R. Petta, A.~C. Johnson, J.~M. Taylor, E.~A. Laird, A.~Yacoby, M.~D. Lukin,
  C.~M. Marcus, M.~P. Hanson, and A.~C. Gossard, Science \textbf{309}, 2180
  (2005).

\bibitem{PhysRevB.59.2070}
G.~Burkard, D.~Loss, and D.~P. DiVincenzo, Phys. Rev. B \textbf{59}, 2070
  (1999).

\bibitem{PhysRevA.90.022320}
A.~Auer and G.~Burkard, Phys. Rev. A \textbf{90}, 022320 (2014).

\bibitem{PhysRevA.54.3824}
C.~H. Bennett, D.~P. DiVincenzo, J.~A. Smolin, and W.~K. Wootters, Phys. Rev. A
  \textbf{54}, 3824 (1996).

\bibitem{PhysRevA.40.4277}
R.~F. Werner, Phys. Rev. A \textbf{40}, 4277 (1989).

\bibitem{PhysRevA.78.062313}
T.~Tanamoto, K.~Maruyama, Y.-x. Liu, X.~Hu, and F.~Nori, Phys. Rev. A
  \textbf{78}, 062313 (2008).

\bibitem{PhysRevA.78.022312}
K.~Maruyama and F.~Nori, Phys. Rev. A \textbf{78}, 022312 (2008).

\bibitem{PhysRevA.84.042303}
D.~Gon\ifmmode~\mbox{\c{t}}\else \c{t}\fi{}a and P.~van Loock, Phys. Rev. A
  \textbf{84}, 042303 (2011).

\bibitem{PhysRevA.86.052312}
D.~Gon\ifmmode~\mbox{\c{t}}\else \c{t}\fi{}a and P.~van Loock, Phys. Rev. A
  \textbf{86}, 052312 (2012).

\bibitem{PhysRevLett.94.236803}
J.~M. Taylor, W.~D\"ur, P.~Zoller, A.~Yacoby, C.~M. Marcus, and M.~D. Lukin,
  Phys. Rev. Lett. \textbf{94}, 236803 (2005).

\bibitem{Pan:2001}
J.-W. Pan, C.~Simon, C.~Brukner, and A.~Zeilinger, Nature \textbf{410}, 1067
  (2001).

\bibitem{RevModPhys.79.1217}
R.~Hanson, L.~P. Kouwenhoven, J.~R. Petta, S.~Tarucha, and L.~M.~K.
  Vandersypen, Rev. Mod. Phys. \textbf{79}, 1217 (2007).

\bibitem{PhysRevLett.83.4204}
A.~Imamo\ifmmode~\breve{g}\else \u{g}\fi{}lu, D.~D. Awschalom, G.~Burkard,
  D.~P. DiVincenzo, D.~Loss, M.~Sherwin, and A.~Small, Phys. Rev. Lett.
  \textbf{83}, 4204 (1999).

\bibitem{RevModPhys.73.357}
Y.~Makhlin, G.~Sch\"on, and A.~Shnirman, Rev. Mod. Phys. \textbf{73}, 357
  (2001).

\bibitem{Abragam1961}
A.~Abragam, \emph{Principles of Nuclear Magnetism}, Oxford University Press,
  Oxford (1961).

\bibitem{Neumann:2010}
P.~Neumann, R.~Kolesov, B.~Naydenov, J.~Beck, F.~Rempp, M.~Steiner, V.~Jacques,
  G.~Balasubramanian, M.~L. Markham, D.~J. Twitchen, S.~Pezzagna, J.~Meijer,
  J.~Twamley, F.~Jelezko, and J.~Wrachtrup, Nature Phys. \textbf{6}, 249
  (2010).

\bibitem{PhysRevA.63.042111}
T.~Eggeling and R.~F. Werner, Phys. Rev. A \textbf{63}, 042111 (2001).

\bibitem{GAMM:GAMM200890003}
G.~Dirr and U.~Helmke, GAMM-Mitteilungen \textbf{31}, 59 (2008).

\bibitem{Brockett:1973}
R.~Brockett, in \emph{Geometric Methods in System Theory}, volume~3 of
  \emph{NATO Advanced Study Institutes Series} (edited by D.~Mayne and
  R.~Brockett), 43--82, Springer Netherlands (1973).

\bibitem{chinchuluun2007}
A.~Chinchuluun, P.~M. Pardalos, A.~Migdalas, and L.~Pitsoulis (Editors),
  \emph{Pareto Optimality, Game Theory And Equilibria}, Springer, New York
  (2008).

\bibitem{Dzyaloshinsky1958241}
I.~Dzyaloshinsky, Journal of Physics and Chemistry of Solids \textbf{4}, 241
  (1958).

\bibitem{PhysRev.120.91}
T.~Moriya, Phys. Rev. \textbf{120}, 91 (1960).

\end{thebibliography}

\end{document}